\documentclass[sigconf]{acmart}
\usepackage{tabularx}
\usepackage{dsfont}
\usepackage{amsmath}
\usepackage{graphicx}
\usepackage{tabularx}
\usepackage{listings}
\usepackage{subcaption}
\usepackage{multirow}
\usepackage{adjustbox}
\usepackage{tablefootnote}
\usepackage{nidanfloat}

\usepackage{amsmath,amsfonts,bm}

\def\eqref#1{equation~\ref{#1}}

\def\1{\bm{1}}

\def\vs{{\bm{s}}}

\def\vx{{\bm{x}}}
\def\vy{{\bm{y}}}
\def\vz{{\bm{z}}}

\DeclareMathAlphabet{\mathsfit}{\encodingdefault}{\sfdefault}{m}{sl}
\SetMathAlphabet{\mathsfit}{bold}{\encodingdefault}{\sfdefault}{bx}{n}

\AtBeginDocument{%
  \providecommand\BibTeX{{%
    \normalfont B\kern-0.5em{\scshape i\kern-0.25em b}\kern-0.8em\TeX}}}

\copyrightyear{2020}
\acmYear{2010}
\setcopyright{rightsretained}

\setcopyright{rightsretained}
\acmConference[SIGIR eCom'20]{ACM SIGIR Workshop on eCommerce}{July 30, 2020}{Virtual Event, China}
\acmYear{2020}
\copyrightyear{2020}
\makeatletter
\renewcommand\@formatdoi[1]{\ignorespaces}
\makeatother
\acmISBN{}

\begin{document}

\title{Context-Aware Learning to Rank with Self-Attention}

\author{Przemys\l{}aw Pobrotyn}
\affiliation{%
  \institution{ML Research at Allegro.pl}
}
\email{przemyslaw.pobrotyn@allegro.pl}
\author{Tomasz Bartczak}
\affiliation{%
  \institution{ML Research at Allegro.pl}
}
\email{tomasz.bartczak@allegro.pl}
\author{Miko\l{}aj Synowiec}
\affiliation{%
  \institution{ML Research at Allegro.pl}
}
\email{mikolaj.synowiec@allegro.pl}
\author{Rados\l{}aw Bia\l{}obrzeski}
\affiliation{%
  \institution{ML Research at Allegro.pl}
}
\email{radoslaw.bialobrzeski@allegro.pl}

\author{Jaros\l{}aw Bojar}
\affiliation{%
  \institution{ML Research at Allegro.pl}
}
\email{jaroslaw.bojar@allegro.pl}

\renewcommand{\shortauthors}{Pobrotyn et al.}

\begin{abstract}
Learning to rank is a key component of many e-commerce search engines. In learning to rank, one is interested in optimising the global ordering of a list of items according to their utility for users. Popular approaches learn a scoring function that scores items individually (i.e. without the context of other items in the list) by optimising a pointwise, pairwise or listwise loss. The list is then sorted in the descending order of the scores. Possible interactions between items present in the same list are taken into account in the training phase at the loss level. However, during inference, items are scored individually, and possible interactions between them are not considered. In this paper, we propose a context-aware neural network model that learns item scores by applying a self-attention mechanism. The relevance of a given item is thus determined in the context of all other items present in the list, both in training and in inference. We empirically demonstrate significant performance gains of self-attention based neural architecture over Multi-Layer Perceptron baselines, in particular on a dataset coming from search logs of a large scale e-commerce marketplace, Allegro.pl. This effect is consistent across popular pointwise, pairwise and listwise losses. Finally, we report new state-of-the-art results on MSLR-WEB30K, the learning to rank benchmark.
\end{abstract}

\ccsdesc[500]{Information systems}
\ccsdesc[500]{Information systems~Learning to rank}

\keywords{learning to rank, self-attention, context-aware ranking}

\maketitle

\section{Introduction}
Learning to rank (LTR) is an important area of machine learning research, lying at the core of many information retrieval (IR) systems. It arises in numerous industrial applications like e-commerce search engines, recommender systems, question-answering systems, and others.

A typical machine learning solution to the LTR problem involves learning a scoring function, which assigns real-valued scores to each item of a given list, based on a dataset of item features and human-curated or implicit (e.g. clickthrough logs) relevance labels. Items are then sorted in the descending order of scores \cite{Liu:2009:LRI:1618303.1618304}. Performance of the trained scoring function is usually evaluated using an IR metric like Mean Reciprocal Rank (MRR) \cite{MRR}, Normalised Discounted Cumulative Gain (NDCG) \cite{NDCG} or Mean Average Precision (MAP) \cite{MAP}. 

In contrast to other classic machine learning problems like classification or regression, the main goal of a ranking algorithm is to determine relative preference among a group of items. Scoring items individually is a proxy of the actual learning to rank task. Users' preference for a given item on a list depends on other items present in the same list: an otherwise preferable item might become less relevant in the presence of other, more relevant items. For example, in the context of an e-commerce search engine, the relative desirability of an item might depend of the relation of its price to the prices of other items displayed in the results list. Common learning to rank algorithms attempt to model such inter-item dependencies at the loss level. That is, items in a list are still scored individually, but the effect of their interactions on evaluation metrics is accounted for in the loss function, which usually takes a form of a pairwise (RankNet \cite{RankNet}, LambdaLoss \cite{LambdaLoss}) or a listwise (ListNet \cite{ListNet}, ListMLE \cite{ListMLE}) objective. For example, in LambdaMART \cite{burges2010ranknet} the gradient of the pairwise loss is rescaled by the change in NDCG of the list which would occur if a pair of items was swapped. Pointwise objectives, on the other hand, do not take such dependencies into account. 

In this work, we propose a learnable, context-aware, self-attention \cite{attention} based scoring function, which allows for modelling of inter-item dependencies not only at the loss level but also in the computation of items' scores. Self-attention is a mechanism first introduced in the context of natural language processing. Unlike RNNs \cite{Hochreiter:1997:LSM:1246443.1246450}, it does not process the input items sequentially but allows the model to attend to different parts of the input regardless of their distance from the currently processed item. We adapt the Transformer \cite{attention}, a popular self-attention based neural machine translation architecture, to the ranking task. Since the self-attention operation is permutation-equivariant (scores items the same way irrespective of their input order), we obtain a permutation-equivariant scoring function suitable for ranking. If we further refine the model with positional encodings, the obtained model becomes suitable for re-ranking setting. We demonstrate that the obtained (re)ranking model significantly improves performance over Multi-Layer Perceptron (MLP) baselines across a range of pointwise, pairwise and listwise ranking losses. Evaluation is conducted on MSLR-WEB30K \cite{Web30K}, the benchmark LTR dataset with multi-level relevance judgements, as well as on clickthrough data coming from Allegro.pl, a large-scale e-commerce search engine. We also establish the new state-of-the-art results on WEB30K in terms of NDCG@5.

We provide an open-source Pytorch \cite{paszke2017automatic} implementation of our self-attentive context-aware ranker available at \url{https://github.com/allegro/allRank}.

The rest of the paper is organised as follows. In Section \ref{Related work} we review related work. In Section \ref{Problem formulation} we formulate the problem solved in this work. In Section \ref{self attention} we describe our self-attentive ranking model. Experimental results and their discussion are presented in Section \ref{Experiments}. In Section \ref{ablation} we conduct an ablation study of various hyperparameters of our model. Finally, a summary of our work is given in Section \ref{Conclusion}.

\section{Related work}
\label{Related work}
Learning to rank has been extensively studied and there is a plethora of resources available on classic pointwise, pairwise and listwise approaches. We refer the reader to \cite{Liu:2009:LRI:1618303.1618304} for the overview of the most popular methods. 

What the majority of LTR methods have in common is that their scoring functions score items individually. Inter-item dependencies are (if at all) taken into account at the loss level only. Previous attempts at modelling context of other items in a list in the scoring function include:
\begin{itemize}
    \item a pairwise scoring function \cite{Dehghani:2017:NRM:3077136.3080832} and Groupwise Scoring Function (GSF) \cite{GroupwiseScoring}, which incorporates the former work as its special case. However, the proposed GSF method simply concatenates feature vectors of multiple items and passes them through an MLP. To desensitize the model to the order of concatenated items, Monte-Carlo sampling is used, which yields an unscalable algorithm,
    \item a seq2slate model \cite{seq2slate} uses an RNN combined with a variant of Pointer Networks \cite{PointerNetworks} in an encoder-decoder type architecture to both encode items in a context-aware fashion and then produce the optimal list by selecting items one-by-one. Authors evaluate their approach only on clickthrough data (both real and simulated from WEB30K). A similar, simpler approach known as Deep Listwise Context Model (DLCM) was proposed in \cite{DBLP:journals/corr/abs-1804-05936}: an RNN is used to encode a set of items for re-ranking, followed by a single decoding step with attention,
    \item in \cite{saratch2019exploring}, authors attempt to capture inter-item dependencies by adding so-called \textit{delta} features that represent how different a given item is from items surrounding it in the list. It can be seen as a simplified version of a local self-attention mechanism. Authors evaluate their approach on proprietary search logs only, 
    \item authors of \cite{jiang2018beyond} formulate the problem of re-ranking of a list of items as that of a whole-list generation. They introduce ListCVAE, a variant of Conditional Variational Auto-Encoder \cite{NIPS2015_5775} which learns the joint distribution of items in a list conditioned on users' relevance feedback and uses it to directly generate a ranked list of items. Authors claim NDCG unfairly favours greedy ranking methods and thus do not use that metric in their evaluation,   
    \item similarly to our approach, \citet{pei2019personalized} use the self-attention mechanism to model inter-item dependencies. Their approach, however, was not evaluated on a standard WEB30K dataset and the only loss function considered was ListNet.
\end{itemize} 

Our proposed solution to the problem of context-aware ranking makes use of the self-attention mechanism. It was first introduced as intra-attention in \cite{cheng-etal-2016-long} and received more attention after the introduction of the Transformer architecture \cite{attention}. Our model can be seen as a special case of the encoder part of the Transformer.

Our model, being a neural network, can be trained with gradient-based optimisation methods to minimise any differentiable loss function. Loss functions suitable for the ranking setting have been studied extensively in the past \cite{Liu:2009:LRI:1618303.1618304}. In order to demonstrate that our context-aware model provides performance boosts irrespectively of the loss function used, we evaluate its performance when tasked with optimising several popular ranking losses. 

We compare the proposed approach with those of the aforementioned methods which provided an evaluation on WEB30K in terms of NDCG@5 and NDCG@10. These include GSF of \cite{GroupwiseScoring} and DLCM of \cite{DBLP:journals/corr/abs-1804-05936}. We outperform both competing methods.

\section{Problem formulation}
\label{Problem formulation}
In this section, we formulate problem at hand in learning to rank setting. Let $X$ be the training set. It consists of pairs $(\vx, \vy)$ of a list $\vx$ of $d_f$-dimensional real-valued vectors $x_i$ together with a list $\vy$ of their relevance labels $y_i$ (multi-level or binary). Note that lists $\vx$ in the training set may be of varying length. The goal is to find a scoring function $f$ which maximises an IR metric of choice (e.g. NDCG) on the test set. Since IR metrics are rank based (thus, non-differentiable), the scoring function $f$ is trained to minimise the average of a surrogate loss $l$ over the training data.    
$$
\mathcal{L}(f) = \frac{1}{|X|}\sum_{(\vx, \vy)\in X}l((\vx,\vy), f),
$$
while controlling for overfitting (e.g. by using dropout \cite{Srivastava:2014:DSW:2627435.2670313} in the neural network based scoring function $f$ or adding $L_1$ or $L_2$ penalty term \cite{Ng:2004:FSL:1015330.1015435} to the loss function $l$ ). 
Thus, two crucial choices one needs to make when proposing a learning to rank algorithm are that of a scoring function $f$ and loss function $l$. As discussed earlier, typically, $f$ scores elements $x_i \in \vx$ individually to produce scores $f(x_i)$, which are then input to loss function $l$ together with ground truth labels $y_i$. In subsequent sections, we describe our construction of context-aware scoring function $f$ which can model interactions between items $x_i$ in a list $\vx$. Our model is generic enough to be applicable with any of standard pointwise, pairwise or listwise loss. We thus experiment with a variety of popular ranking losses $l$.

\section{Self-attentive ranker}
\label{self attention}
In this section, we describe the architecture of our self-attention based ranking model. We modify the Transformer architecture to work in the ranking setting and obtain a scoring function which, when scoring a single item, takes into account all other items present in the same list.

\subsection{Self-Attention Mechanism}
The key component of our model is the self-attention mechanism introduced in \cite{attention}.
The attention mechanism can be described as taking the query vector and pairs of key and value vectors as input and producing a vector output. The output of the attention mechanism for a given query is a weighted sum of the value vectors, where weights represent how relevant to the query is the key of the corresponding value vector. Self-attention is a variant of attention in which query, key and value vectors are all the same - in our case, they are vector representations of items in the list. The goal of the self-attention mechanism is to compute a new, higher-level representation for each item in a list, by taking a weighted sum over all items in a list according to weights representing the relevance of these items to the query item.

There are many ways in which one may compute the relevance of key vectors to query vectors. We use the variant of self-attention known as \textit{Scaled Dot-Product Attention}. Suppose $Q$ is a $d_{\mathrm{model}}$\nobreakdash-dimensional matrix representing all items (queries) in the list. Let $K$ and $V$ be the keys and values matrices, respectively. Then
$$
\mathrm{Attention}(Q, K, V) = \mathrm{softmax}(\frac{QK^T}{\sqrt{d_{\mathrm{model}}}})V.
$$
The scaling factor of $\frac{1}{\sqrt{d_{\mathrm{model}}}}$ is added to avoid small gradients in the softmax operation for large values of $d_{\mathrm{model}}$. 

\subsection{Multi-Headed Self-Attention}
As described in \cite{attention}, it is beneficial to perform the self-attention operation multiple times and concatenate the outputs. To avoid growing the size of the resulting output vector, matrices $Q$, $K$ and~$V$ are first linearly projected $H$ times to $d_q$, $d_k$ and $d_v$ dimensional spaces, respectively. Usually, $d_q = d_k = d_v = d_{\mathrm{model}} / H$. Each of $H$ computations of a linear projection of $Q$, $K$, $V$, followed by a self-attention mechanism is referred to as a single attention head. Note that each head has its own learnable projection matrices. The outputs of each head are concatenated and once again linearly projected, usually to the vector space of the same dimension as that of input matrix $Q$. Similarly to the Transformer, our model also uses multiple attention heads.
Thus
$$
\mathrm{MultiHead(Q, K, V)} = \mathrm{Concat(head_1, \ldots, head_H)}W^O
$$
$$
\mathrm{where}\text{  }\mathrm{head}_i = \mathrm{Attention}(QW_i^Q, KW_i^K, VW_i^V)
$$
and the projections are given by matrices 
$$W_i^Q \in \mathbb{R}^{d_{model} \times d_q}, \text{  } W_i^K \in \mathbb{R}^{d_{model} \times d_k}, $$
$$W_i^V \in \mathbb{R}^{d_{model} \times d_v}, \text{  } W^O \in \mathbb{R}^{Hd_v \times d_{model}}.$$

\subsection{Permutation-equivariance}
The key property of the proposed context-aware model making it suitable for the ranking setting is that it is permutation-equivariant, i.e. the scores of items do not depend on the original ordering of the input. Recall the definition of permutation equivariance:

\theoremstyle{definition}
\begin{definition}
Let $\vx \in \mathbb{R}^n$ be a real-valued vector and $\pi \in S_n$ be a permutation of $n$ elements. A function $f: \mathbb{R}^n \rightarrow \mathbb{R}^n$ is called permutation-equivariant iff
$$
f(\pi(\vx)) = \pi(f(\vx)).
$$

That is, a function is permutation-equivariant if it commutes with any permutation of the input elements.

It is a trivial observation that the self-attention operation is permutation-equivariant.

\end{definition}

\subsection{Positional Encodings}
Transformer architecture was designed to solve a neural machine translation (NMT) task. In NMT, the order of input tokens should be taken into account. Unlike RNNs, self-attention based encoder has no way of discerning the order of input tokens, because of its permutation-equivariance. Authors of the original Transformer paper proposed to solve the problem by adding either fixed or learnable positional encodings to the input embeddings. Fixed positional encodings use sine and cosine functions of different frequencies, as follows:
$$
PE_{(pos, 2i)} = \sin{(pos / 10000^{2i/d_{model}})}
$$

$$
PE_{(pos, 2i + 1)} = \cos{(pos / 10000^{2i/d_{model}})}
$$
where $pos$ is the position and $i$ is the dimension.

The ranking problem can be viewed as either ordering a set of (unordered) items or as re-ranking, where the input list has already been sorted according to a weak ranking model. In the former case, the use of positional encodings is not needed. In the latter, they may boost the model's performance. We experiment with both ranking and re-ranking settings and when positional encodings are used, we test the fixed encodings variant \footnote{We found learnable positional encodings to yield similar results.}. Details can be found in Section \ref{Experiments}. 

\subsection{Model Architecture}
\begin{figure}[h]
    \centering
    \includegraphics[width=1.\linewidth]{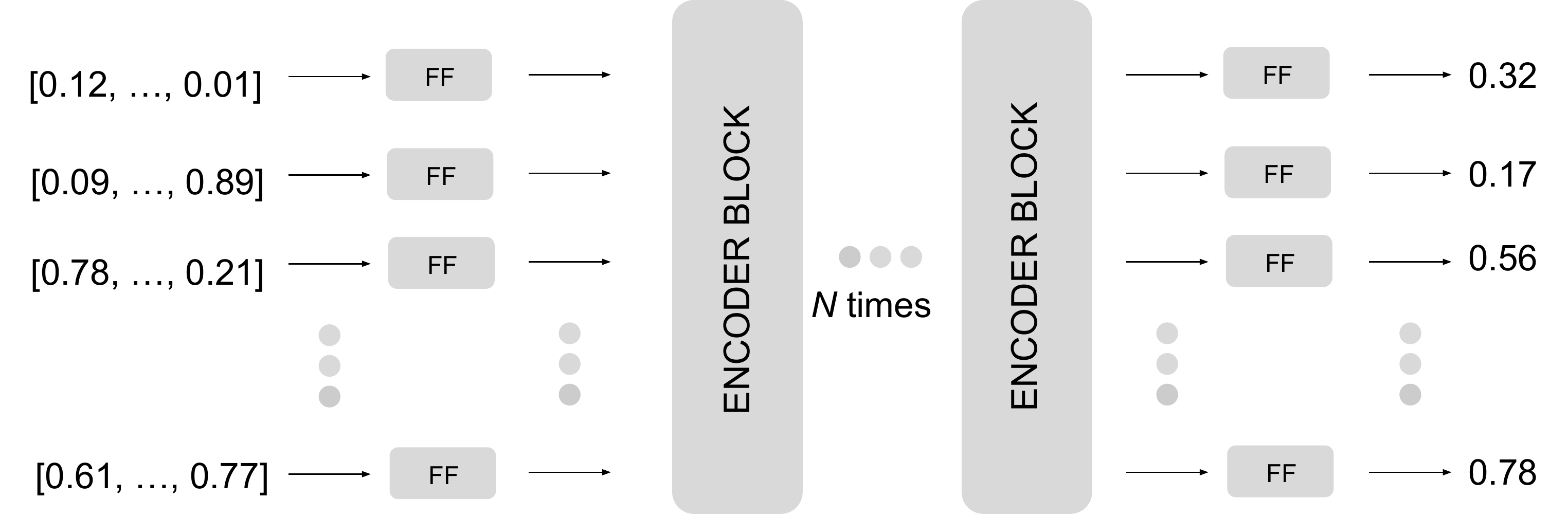}
    \caption{Schematic of the proposed model architecture. Input is a list of real-valued vectors. Output is the list of real-valued scores.}
    \label{fig:architecture_schematic}
\end{figure}

We adapt the Transformer model to the ranking setting as follows. Items on a list are treated as tokens and item features as input token embeddings. We denote the length of an input list as $l$ and the number of features as $d_f$. Each item is first passed through a shared fully connected layer of size $d_{fc}$. Next, hidden representations are passed through an encoder part of Transformer architecture with $N$ encoder blocks, $H$ heads and hidden dimension $d_h$. Recall that an encoder block in the Transformer consists of a multi-head attention layer with a skip-connection \cite{He2015DeepRL} to the input, followed by layer normalisation \cite{Ba2016LayerN}, time-distributed feed-forward layer, and another skip connection followed by layer normalisation. Dropout is applied before performing summation in residual blocks.
Finally, after $N$ encoder blocks, a fully-connected layer shared across all items in the list is used to compute a score for each item. The model can be seen as an encoder part of the Transformer with extra linear projection on the input (see Figure \ref{fig:architecture_schematic} for a schematic of the architecture).

Thus, the model can be expressed as 
$$
f(\vx) = \mathrm{FC}(\underbrace{\mathrm{Encoder}(\mathrm{Encoder}(...(\mathrm{Encoder}}_{N \text{times}}(\mathrm{FC}(\vx))))))
$$

where
$$
\mathrm{Encoder}(\vx) = \mathrm{LayerNorm}(\vz + \mathrm{Dropout}(\mathrm{FC}(\vz))),
$$
$$
\vz = \mathrm{LayerNorm}(\vx + \mathrm{Dropout}(\mathrm{MultiHead}(\vx)))
$$

and $\mathrm{FC}$, $\mathrm{Dropout}$ are fully-connected and dropout layers, respectively.

By using self-attention in the encoder, we ensure that in the computation of a score of a given item, hidden representation of all other items were accounted for. Obtained scores, together with ground truth labels, can provide input to any ranking loss of choice. If the loss is a differentiable function of scores (and thus, of model's parameters), one can use SGD to optimise it. We thus obtain a general, context-aware model for scoring items on a list that can readily be used with any differentiable ranking loss. Since all the components used in the construction of the model (self-attention, layer normalisation, feed-forward layers) are permutation-equivariant, the entire model is permutation-equivariant (unless positional encodings are used).

\section{Experiments}
\label{Experiments}
\subsection{Datasets}
Learning to rank datasets come in two flavours: they can have either multi-level or binary relevance labels. Usually, multi-level relevance labels are human-curated, whereas binary labels are derived from clickthrough logs and are considered implicit feedback. We evaluate our context-aware ranker on both types of data. 

For the first type, we use the popular WEB30K dataset, which consists of more than 30,000 queries together with lists of associated search results. Every search result is encoded as a $136$-dimensional real-valued vector and has associated with it a relevance label on the scale from 0 (irrelevant) to 4 (most relevant). We standardise the features before inputting them into a learning algorithm. The dataset comes partitioned into five folds with roughly the same number of queries per fold. We perform 5-fold cross-validation by training our models on three folds, validating on one and testing on the final fold. All results reported are averages across five folds together with the standard deviation of results. Since lists in the dataset are of unequal length, we pad or subsample to equal length for training, but use full length (i.e. pad to maximum length present in the dataset) for validation and testing. Note that there are 982 queries for which the associated search results list contains no relevant documents (i.e. all documents have label 0). For such lists, the NDCG can be arbitrarily set to either 0 or 1. To allow for a fair comparison with the current state-of-the-art, we followed LightGBM \cite{lightbGBM} implementation of setting NDCG of such lists to 1 during evaluation.

For a dataset with binary labels, we use clickthrough logs of a large scale e-commerce search engine from Allegro.pl. The search engine already has a ranking model deployed, which is trained using XGBoost \cite{Chen:2016:XST:2939672.2939785} with \texttt{rank:pairwise} loss. We thus treat learning on this dataset as a re-ranking problem and use fixed positional encodings in context-aware scoring functions. This lets the models leverage items' positions returned by the base ranker. The search logs consist of $1$M lists, each of length at most 60. Nearly all lists (95\%) have only one relevant item with label 1; remaining items were not clicked and are deemed irrelevant (label 0). Each item in a list is represented by a 45-dimensional, real-valued vector. We do not perform cross-validation on this set, but we use the usual train, validation and test splits of the data (using 100k lists for validation, 100k for test and the remaining lists for training).
\subsection{Loss Functions}
\label{Losses}

To evaluate the performance of the proposed context-aware ranking model, we use several popular ranking losses. Pointwise losses used are RMSE of predicted scores and ordinal loss \cite{OrdinalRegression} (with minor modification to make it suitable for ranking). For pairwise losses, we use NDCGLoss 2++ (one of the losses of LambdaLoss framework) and its special cases, RankNet and LambdaRank \cite{LambdaRank}. Listwise losses used consist of ListNet and ListMLE.

Below, we briefly describe all of the losses used. For a more thorough treatment, please refer to the original papers. Throughout, $X$ denotes the training set, $\vx$ denotes an input list of items, $\vs = f(\vx)$ is a vector of scores obtained via the ranking function $f$ and $\vy$ is the vector of ground truth relevancy labels.

\subsubsection{Pointwise RMSE}
The simplest baseline is a pointwise loss, in which no interaction between items is taken into account. We use RMSE loss:
$$
l(\vs, \vy) = \sqrt{\sum_i(y_i - s_i)^2}
$$
In practice, we used sigmoid activation function on the outputs of the scoring function $f$ and rescaled them by multiplying by maximum relevance value (e.g. 4 for WEB30K).

\subsubsection{Ordinal Loss}
We formulated ordinal loss as follows. Multi-level ground truth labels were converted to vectors as follows:
\begin{align*}
 0 \mapsto [0, 0, 0 ,0],\\
1 \mapsto [1, 0, 0, 0],\\
2 \mapsto [1, 1, 0, 0],\\
3 \mapsto [1, 1, 1, 0],\\
4 \mapsto [1, 1, 1, 1]. 
\end{align*}

The self-attentive scoring function was modified to return four outputs and each output was passed through a sigmoid activation function. Thus, each neuron of the output predicts a single relevancy level, but by the reformulation of ground truth, their relative order is maintained, i.e. if, say, label 2 is predicted, label 1 should be predicted as well (although it is not strictly enforced and model is allowed to predict label 2 without predicting label 1). The final loss value is the mean of binary cross-entropy losses for each relevancy level. During inference, the outputs of all output neurons are summed to produce the final score of an item. Note that this is the classic ordinal loss, simply used in the ranking setting. 

\subsubsection{LambdaLoss, RankNet and LambdaRank}
We used NDCG-Loss2++ of \cite{LambdaLoss}, formulated as follows:
$$
l(\vs, \vy) = -\sum_{y_i > y_j} \log_2 \sum_{\pi}\left(\frac{1}{1+e^{-\sigma(s_i - s_j)}}\right)^{(\rho_{ij} + \mu\delta_{ij})|G_i -G_j|}H(\pi|\vs) 
$$
where 
$$
G_i = \frac{2^{y_i} - 1}{\mathrm{maxDCG}},
$$
$$
\rho_{ij} = \left|\frac{1}{D_i} - \frac{1}{D_j}\right|,
$$
$$
\delta_{ij} = \left|\frac{1}{D_{|i-j|}} - \frac{1}{D_{|i-j|} +1}\right|,
$$
$$
D_i = log_2(1+i)
$$
and $H(\pi|\vs)$ is a hard assignment distribution of permutations, i.e. 
$$
H(\hat{\pi}|\vs) = 1 \text{ and } H(\pi|\vs) = 0 \text{ for all } \pi \neq \hat{\pi}
$$
where $\hat{\pi}$ is the permutation in which all items are sorted by decreasing scores $\vs$. Fixed parameter $\mu$ is set to $10.0$.

By removing the exponent in $l(\vs, \vy)$ formula we obtain the RankNet loss function, weighing each score pair identically. Similarly, we may obtain differently weighted RankNet variants by changing the formula in the exponent.

To obtain a LambdaRank formula, replace the exponent with
$$
\Delta\mathrm{NDCG}(i, j) = |G_i - G_j|\rho_{ij}.
$$

\subsubsection{ListNet and ListMLE}
ListNet loss \citep{ListNet} is given by the following formula:
$$
l(\vs, \vy) = - \sum_j \mathrm{softmax}(\vy)_j \times \log(\mathrm{softmax}(\vs)_j)
$$
In binary version, softmax of ground truth $\vy$ is omitted for single-click lists and replaced with normalisation by the number of clicks for multiple-click lists. 

ListMLE \citep{ListMLE} is given by:
$$
l(\vs, \vy) = - \log P(\vy | \vs)
$$
where
$$
P(\vy | \vs) = \prod_i^{n}\frac{\exp(f(x_{y(i)}))}{\sum_{k=i}^{n}\exp(f(x_{y(k)}))}
$$
and $y(i)$ is the index of object which is ranked at position $i$.

\begin{table}[h]
\caption{Results from the literature on WEB30K}
\label{other-results}
\begin{center}
\begin{tabular}{l|ccc}
\multicolumn{1}{c|}{\bf Method}  
&\multicolumn{1}{c}{\bf NDCG@5} 
&\multicolumn{1}{c}{\bf NDCG@10} \\
\hline
GSF \footnotemark &  44.46  & 46.77\\
DLCM & 45.00 & 46.90 \\
NDCGLoss 2++ (LightGBM) & 51.21 & - \\
Context-Aware Ranker (this work) & \textbf{53.00} & \textbf{54.88}\\
\hline
\end{tabular}
\end{center}
\end{table}
\footnotetext{Please note that the GSF result was calculated after dropping the approx. $3\%$ queries without any relevant documents, as reported in \cite{GroupwiseScoring}.}

\subsection{Experimental setup}
We train both our context-aware ranking models and MLP models on both datasets, using all loss functions discussed in Section \ref{Losses}~\footnote{For the clickthrough logs dataset, we used only the losses which can be applied to binary relevance labels.}. We also train XGBoost models with \texttt{rank:pairwise} loss similar to the production model of the e-commerce search engine for both datasets. Hyperparameters of all models (number of encoder blocks, number of attention heads, dropout, etc.) are tuned on the validation set of Fold 1 for each loss separately. MLP models are constructed to have a similar number of parameters to context-aware ranking models. For optimisation of neural network models, we use Adam optimiser \citep{Adam} with the learning rate tuned separately for each model. Details of hyperparameters used can be found in Appendix~\ref{appendix_hyperparams}. In Section \ref{ablation} we provide an ablation study of the effect of various hyperparameters on the model's performance.

\begin{table}[h]
    \caption{Relative percentage NDCG@60 improvement on e-commerce search logs dataset}
    \label{main-results-binary}
    \begin{tabular}{l|cc}
    \multicolumn{1}{c|}{\bf Loss}  
    &\multicolumn{1}{c}{\bf Self-attention}
    &\multicolumn{1}{c}{\bf MLP} \\
    \hline
     NDCGLoss 2++ &  \textbf{3.00} &   \textbf{1.51}\\
         LambdaRank &  2.97 &  1.39 \\
            ListNet &  2.93 &  1.24 \\
            RankNet &  2.68 &  1.19 \\
    \hline
        & \multicolumn{2}{c}{\bf XGBoost} \\
        \hline
        \texttt{rank:pairwise} & \multicolumn{2}{c}{1.83} \\
    \hline
    \end{tabular}
\end{table}

\subsection{Results}

On WEB30K, models' performance is evaluated using NDCG@5\footnote{Expressed as a percentage.},  which is the usual metric reported for this dataset, as well as NDCG at rank cutoffs 10 and 30. Results are reported in Table \ref{main-results}. On e-commerce search logs, we report a relative percentage increase in NDCG@60 \footnote{We chose 60 as users are presented with search results lists of length 60.} over production XGBoost model, presented in Table \ref{main-results-binary}. We observe consistent and significant performance improvement of the proposed self-attention based model over MLP baseline across all types of loss functions considered and any of the chosen rank cutoffs. In particular, for ListNet we observe a $7.3\%$ performance improvement over MLP baseline on WEB30K in terms of NDCG@5. Note also that the best performing MLP model is outperformed even by the worst-performing self-attention based model on both datasets in all the metrics reported. We thus observe that incorporating context-awareness into the model architecture has a more pronounced effect on the performance of the model than varying the underlying loss function. Surprisingly, ordinal loss outperforms more established and better-studied losses like ListNet, ListMLE or NDCGLoss 2++ on multi-level relevancy data. In particular, we improve on the previous state-of-the-art NDCG@5 result of 51.21 by 2.27\%, obtaining 52.37. The previous state-of-the-art result wasobtained using NDCGLoss 2++ trained using LightGBM. Another surprising finding is a good performance of models trained with RMSE loss, especially as compared to models trained to optimise RankNet and ListMLE. For comparison with the other methods, we provide results on WEB30K reported in other works in Table \ref{other-results}. For models with multiple variants, we cite the best result reported in the original work. In all tables, boldface is the best value column-wise.

\begin{table*}[]
\caption{Test results on WEB30K}
\label{main-results}
\begin{tabular}{l|llllll}
\bf Loss          & \multicolumn{3}{c}{\bf Self-attention}                            & \multicolumn{3}{c}{\bf MLP}              \\ \hline
              & NDCG@5               & NDCG@10 & \multicolumn{1}{l|}{NDCG@30} & NDCG@5 & NDCG@10      & NDCG@30      \\ \hline
Ordinal loss  & \textbf{53.00$\pm$0.35} & \textbf{54.88$\pm$0.21} & \multicolumn{1}{l|}{\bf 60.19$\pm$0.13}       & 48.84$\pm$0.42      & 51.02$\pm$0.33            & 56.98$\pm$0.15            \\
NDCGLoss 2++  & 52.65$\pm$0.37 & 54.49$\pm$0.27 & \multicolumn{1}{l|}{59.80$\pm$0.08}       & \textbf{49.15$\pm$0.44} & \textbf{51.22$\pm$0.34}   & \textbf{57.14$\pm$0.23}           \\
ListNet       & 52.33$\pm$0.33  & 54.26$\pm$0.20 & \multicolumn{1}{l|}{59.63$\pm$0.14}      & 47.81$\pm$0.36     & 50.20$\pm$0.26           & 56.41$\pm$0.17           \\
LambdaRank    & 52.29$\pm$0.31  & 54.08$\pm$0.19 & \multicolumn{1}{l|}{59.48$\pm$0.12}      & 48.77$\pm$0.38 & 50.85$\pm$0.28   & 56.72$\pm$0.17           \\
RMSE          & 51.74$\pm$0.48  & 53.40$\pm$0.38      & \multicolumn{1}{l|}{58.89$\pm$0.31}      & 48.24$\pm$0.52     & 50.31$\pm$0.41           & 56.29$\pm$0.19           \\
RankNet       & 50.79$\pm$0.48  & 52.90$\pm$0.39      & \multicolumn{1}{l|}{58.75$\pm$0.23}     & 47.54$\pm$0.47     & 49.78$\pm$0.35   & 55.91$\pm$0.17  \\
ListMLE       & 50.20$\pm$0.40  & 52.19$\pm$0.23      & \multicolumn{1}{l|}{57.94$\pm$0.18}      & 46.98$\pm$0.45     & 49.14$\pm$0.36           & 55.24$\pm$0.2          \\ \hline
              & \multicolumn{1}{c}{} &         & \multicolumn{2}{c}{\bf XGBoost}           &              &              \\ \cline{2-7} 
              & \multicolumn{2}{c}{NDCG@5}     & \multicolumn{2}{c}{NDCG@10}           & \multicolumn{2}{c}{NDCG@30} \\ \hline
\texttt{rank:pairwise} & \multicolumn{2}{c}{46.8}         & \multicolumn{2}{c}{49.17}                & \multicolumn{2}{c}{55.33}     
\end{tabular}
\end{table*}

\subsection{Re-ranking}
All experiments on WEB30K described above were conducted in the ranking setting - input lists of items were treated as unordered, thus positional encoding was not used. To verify the effect of positional encoding on the model's performance, we conduct the following experiments on WEB30K. To avoid information leak, training data\footnote{In experiments with positional encoding, we used only Fold 1 of the dataset.} is divided into five folds and five XGBoost models are trained, each on four folds. Each model predicts scores for the remaining fold, and the entire dataset is sorted according to these scores.

Finally, we train the same models\footnote{The only loss for which we had to modify the model's hyperparameters as compared to training on the ranking task was NDCGLoss 2++. We observed severe overfitting when training on re-ranking task and thus reduced the number of encoder blocks $N$ from 4 to 2 and hidden dimension $d_h$ form 512 to 256.} as earlier on the sorted dataset but use fixed positional encoding. Results are presented in Table~\ref{results-pe}. For brevity, we focus on NDCG@5. As expected, the models are able to learn positional information and demonstrate improved performance over the plain ranking setting. 

\begin{table}[h]
\caption{NDCG@5 on re-ranking task}
\label{results-pe}
\begin{center}
\begin{tabular}{l|cc}
\multicolumn{1}{c|}{\bf Loss}  
&\multicolumn{1}{c}{\bf With PE}
&\multicolumn{1}{c}{\bf Self-attention w/o PE} \\
\hline
  Ordinal loss &  \textbf{52.67} &           \textbf{52.20} \\
 NDCGLoss 2++ &  52.24 &           51.40 \\
           RMSE &  51.85 &           50.23 \\
        ListNet &  51.77 &           51.34 \\
     LambdaRank &  51.51 &           51.22 \\
        ListMLE &  50.90 &           49.19 \\
        RankNet &  50.58 &           49.90 \\
\hline
\end{tabular}
\end{center}
\end{table}

\subsection{Usage in latency-sensitive conditions}

The results discussed so far were concerned with the \textit{offline} evaluation of the proposed model. Its favourable performance suggests an \textit{online} test in a live search engine, say that of Allegro.pl, is justified. Indeed such a test is planned for further study. It needs to be taken into account though that the complexity of the proposed model is $\mathcal{O}(n^2)$, due to the usage of the self-attention operation.  The inference speed is further influenced by the number $N$ of encoder blocks used. To allow the usage of the model in latency-sensitive scenarios, one might consider using it as a final stage ranker on shorter lists of input elements, reducing the number of encoder blocks, or applying one or more of model distillation \cite{distilling}, quantisation \cite{quantisation} or pruning \cite{pruning} techniques known from the literature.

\begin{table}[h]
\caption{Ablation study}
\label{ablation_study}
\begin{center}
\begin{tabular}{l|cccc}
\multicolumn{1}{c|}{\bf Parameter}  
&\multicolumn{1}{c}{\bf Value}
&\multicolumn{1}{c}{\bf Params} 
&\multicolumn{1}{c}{\bf WEB30K NDCG@5} \\
\hline
baseline &    &      950K &             52.64 \\
\hline
\multirow{2}{*}[1pt]{$H$} &          1 &          950K &               52.44 \\
                        &          4 &          950K &               52.59 \\
\hline
\multirow{2}{*}[1pt]{$N$} &          1 &      250K &               50.62 \\
                        &          2 &      490K &               52.15 \\
\hline
\multirow{4}{*}[1pt]{$d_h$} &         64 &      430K &               51.58 \\
                          &        128 &      509K &               51.96 \\
                          &        256 &      650K &               52.19 \\
                          &       1024 &     1540K &               \textbf{52.86} \\
\hline
\multirow{5}{*}[1pt]{$p_{\mathrm{drop}}$} &        0.0 &          950K &               42.18 \\
                                        &        0.1 &          950K &               51.22 \\
                                        &        0.2 &          950K &               52.58 \\
                                        &        0.3 &          950K &               \textbf{52.86} \\
                                        &        0.5 &          950K &               52.26 \\
\hline
\multirow{4}{*}[1pt]{$l$} &         30 &          950K &               50.94 \\
                        &         60 &          950K &               51.78 \\
                        &        120 &          950K &               52.78 \\
                        &        360 &       950K    &               52.58 \\
\hline
\end{tabular}
\end{center}
\end{table}

\section{Ablation Study}
\label{ablation}
To gauge the effect of various hyperparameters of self-attention based ranker on its performance, we performed the following ablation study. We trained the context-aware ranker with the ordinal loss on Fold 1 of WEB30K dataset, evaluating the ranker on the validation subset. We experimented with a different number $N$ of encoder blocks, $H$ attention heads, length $l$ of longest list used in training, dropout rate $p_{drop}$ and size $d_{h}$ of hidden dimension. Results are summarised in Table \ref{ablation_study}. Baseline model (i.e. the best performing context-aware ranker trained with ordinal loss) had the following values of hyperparameters: $N = 4$, $H = 2$, $l = 240$, $p_{drop} = 0.4$ and $d_{h} = 512$. We observe that a high value of dropout is essential to prevent overfitting but setting it too high results in performance degradation. A similar statement is true of the number of the attention heads - even though it is better to use multiple attention heads as opposed to a single one, we notice a decline in performance when using more than two attention heads. Finally, stacking multiple encoder blocks and increasing the hidden dimension size increases performance. However, we did not test the effect of stacking more than 4 encoder blocks or using hidden dimension sizes larger than 1024 due to GPU memory constraints.  

\section{Conclusions}
\label{Conclusion}
In this work, we addressed the problem of constructing a context-aware scoring function for learning to rank. We adapted the self-attention based Transformer architecture from the neural machine translation literature to propose a new type of scoring function for LTR. We demonstrated considerable performance gains of proposed neural architecture over MLP baselines across different losses and types of data, both in ranking and re-ranking setting. In particular, we established the new state-of-the-art performance on WEB30K. These experiments provide strong evidence that the gains are due to the ability of the model to score items simultaneously. As a result of our empirical study, we observed the strong performance of models trained to optimise ordinal loss function. Such models outperformed models trained with well-studied losses like NDCGLoss 2++ or LambdaRank, which were previously shown to provide tight bounds on IR metrics like NDCG. On the other hand, we observed the surprisingly poor performance of models trained to optimise RankNet and ListMLE losses. In future work, we plan to investigate the reasons for both good and poor performance of the aforementioned losses, in particular, the relation between ordinal loss and NDCG.


\bibliographystyle{ACM-Reference-Format}
\bibliography{bibliography}

\appendix
\section{Experimental details}
\label{appendix_hyperparams}
In Tables \ref{experiment_details_table} and \ref{experiment_details_table_mlp} we provide hyperparameters used for all models reported in Table \ref{main-results}. Models trained on WEB30K were trained for 100 epochs with the learning rate decayed by $0.1$ halfway through the training. On e-commerce search logs, we trained the models for 10 epochs and decayed the learning rate by $0.1$ after $5$-th epoch.  The meaning of the columns in Table \ref{experiment_details_table} is as follows: $d_{f_c}$ is the dimension of the linear projection done on the input data before passing it to the context-aware ranker, $N$ is the number of encoder blocks, $H$ is the number of attention heads, $d_h$ is the hidden dimension used throughout computations in encoder blocks, \textit{lr} is the learning rate, $p_{\mathrm{drop}}$ is the dropout probability and $l$ is the list length (lists of items were either padded or subsampled to that length). The last column shows the number of learnable parameters of the model.

In Table \ref{experiment_details_table_mlp}, \textit{Hidden dimensions} column gives dimensions of subsequent layers of MLP models. The remaining columns have the same meaning as in the other table.

\begin{table*}[b]
\caption{Details of hyperparameters used in self-attentive models}
\label{experiment_details_table}
\begin{center}
\begin{adjustbox}{width=1.0\textwidth}
\begin{tabularx}{\textwidth}{l|@{\hskip 0.1in} *6{>{\centering\arraybackslash}X}cc@{}}
{\bf Loss}
&{$d_{fc}$}
&{$N$}
&{$H$}  
&{$d_{\mathrm{h}}$}
&{$\mathrm{lr}$}
&{$p_{\mathrm{drop}}$}
&{$l$}
&{$|\mathrm{params}|$}\\
\hline
\bf{WEB30K} & & & & & & & &\\
NDCGLoss 2++ & 128 & 4 & 4 & 512 & 1e-3 & 0.3 & 240 & 811K\\
LambdaRank & 128 & 4 & 4 & 512 & 1e-3 & 0.3 & 240 & 811K\\
ListMLE & 256 & 4 & 4 & 512 & 1e-3 & 0.3 & 240 & 2.14M\\
ListNet & 128 & 4 & 4 & 512 & 1e-3 & 0.3 & 240 & 811K\\
Ordinal loss & 144 & 4 & 2 & 512 & 1e-3 & 0.4 & 240 & 949K\\
RankNet & 144 & 4 & 2 & 512 & 1e-3 & 0.3 & 240 & 949K\\
\hline
\bf{E-commerce} & & & & & & & &\\
NDCGLoss 2++ & 128 & 2 & 2 & 128 & 1e-3 & 0.0 & 60 & 206K\\
LambdaRank & 128 & 2 & 2 & 128 & 1e-3 & 0.0 & 60 & 206K\\
ListNet & 128 & 2 & 2 & 128 & 2e-3 & 0.0 & 60 & 206K\\
RankNet & 128 & 2 & 2 & 128 & 2e-3 & 0.0 & 60 & 206K\\
\hline
\end{tabularx}
\end{adjustbox}
\end{center}
\end{table*}

\begin{table*}[b]
\caption{Details of hyperparameters used in MLP models}
\label{experiment_details_table_mlp}
\begin{center}
\begin{adjustbox}{width=1.0\textwidth}
\begin{tabularx}{\textwidth}{l|@{\hskip 0.1in} c *3{>{\centering\arraybackslash}X}c@{}}
{\bf Loss}
&{Hidden dimensions}
&{$\mathrm{lr}$}
&{$p_{\mathrm{drop}}$}
&{$l$}
&{$|\mathrm{params}|$} \\
\hline
\bf{WEB30K} & & & & &\\
all losses & [256, 512, 1024, 512, 256] & 1e-3 & 0.3 & 240 & 1.35M\\
\hline
\bf{E-commerce} & & & & &\\
NDCGLoss 2++ & [256, 384, 256] & 1e-3 & 0.0 & 60 & 210K\\
LambdaRank & [256, 384, 256] & 1e-3 & 0.0 & 60 & 210K\\
ListNet & [256, 384, 256] & 1e-4 & 0.0 & 60 & 210K\\
RankNet & [256, 384, 256] & 1e-4 & 0.0 & 60 & 210K\\
\hline
\end{tabularx}
\end{adjustbox}
\end{center}
\end{table*}
\end{document}